\documentclass[a4paper, superscriptaddress, sd, amsfonts, amssymb, amsmath, preprint, showkeys, nofootinbib, nobibnotes]{revtex4-1}

\usepackage[english]{babel}
\usepackage[utf8]{inputenc}

\usepackage{xcolor}
\usepackage{graphicx}% Include figure files
\usepackage{dcolumn}% Align table columns on decimal point
\usepackage{bm}% bold math
\usepackage{nicefrac}
\usepackage{units}
\usepackage{textcomp}
\usepackage[]{natbib} % Use the natbib reference package - read up on this to edit the reference style; if you want text (e.g. Smith et al., 2012) for the in-text references (instead of numbers), remove 'numbers'
\usepackage{hyperref}
\hypersetup{colorlinks=true, citecolor=blue, urlcolor=cyan} %, citecolor=green, urlcolor=red, linkcolor=blue}
\usepackage{cleveref}
\usepackage{verbatim}
\crefname{section}{§}{§§}
\Crefname{section}{§}{§§}
\usepackage{soul}
\usepackage[normalem]{ulem}

\begin{document}

\title[Inertial focusing]{Inertial focusing of a dilute suspension in pipe flow}
% Force line breaks with \\

\author{Othmane Aouane}
\email{o.aouane@fz-juelich.de}
\affiliation{Helmholtz Institute Erlangen-N\"{u}rnberg for Renewable Energy, Forschungszentrum J\"{u}lich, Cauerstraße 1, 91058 Erlangen, Germany}

\author{Marcello Sega}
\affiliation{Helmholtz Institute Erlangen-N\"{u}rnberg for Renewable Energy, Forschungszentrum J\"{u}lich, Cauerstraße 1, 91058 Erlangen, Germany}

\author{Bastian B\"{a}uerlein}
\affiliation{University of Bremen, Faculty of Production Engineering, Badgasteiner Straße 1, 28359, Bremen, Germany}
\affiliation{Leibniz Institute for Materials Engineering IWT, Badgasteiner Straße 3, 28359, Bremen, Germany}

\author{Kerstin Avila}
\affiliation{University of Bremen, Faculty of Production Engineering, Badgasteiner Straße 1, 28359, Bremen, Germany}
\affiliation{Leibniz Institute for Materials Engineering IWT, Badgasteiner Straße 3, 28359, Bremen, Germany}

\author{Jens Harting}
\affiliation{Helmholtz Institute Erlangen-N\"{u}rnberg for Renewable Energy, Forschungszentrum J\"{u}lich, Cauerstraße 1, 91058 Erlangen, Germany}
\affiliation{Department of Chemical and Biological Engineering and Department of Physics, Friedrich-Alexander-Universit\"at Erlangen-N\"urnberg, Cauerstraße 1, 91058 Erlangen, Germany}

\date{\today}% It is always \today, today,

%%%%% Begin Abstract %%%%%%%%%%%
\begin{abstract}
The dynamics of rigid particle suspensions in a wall-bounded laminar flow present several nontrivial and intriguing features, including particle ordering, lateral transport, and the appearance of stable, preferential locations like the Segr\'{e}-Silberberg annulus. The formation of more than one annulus is a particularly puzzling phenomenon that is still not fully explained. Here, we present numerical simulation results of a dilute suspension of particles in (periodic) pipe flow based on the lattice Boltzmann and the discrete element methods (DEM). Our simulations provide access to the full radial position history of the particles while traveling downstream. This allows to accurately quantify the transient and steady states. We observe the formation of the secondary, inner annulus and show that its position invariably shifts toward the Segr\'{e}-Silberberg one if the channel is sufficiently long, proving that it is, in fact, a transient feature for Reynolds numbers ($Re$) up to $600$. We quantify the variation of the channel focusing length ($L_s/2R$) with $Re$. Interestingly and unlike the theoretical prediction for a point-like particle, we observe that $L_s/2R$ increases with $Re$ for both the single particle and the suspension.
\end{abstract}

\maketitle

\section{Introduction}
\label{sec:introduction}

The idea behind inertial microfluidics stems from the pioneering work of Segr\'{e} \& Silberberg \cite{segre1961,segre1962jfm2} on inertial focusing of neutrally buoyant particles subject to a Poiseuille flow in a straight pipe. Segr\'{e} \& Silberberg observed that the particles migrate across the streamlines and focus on a thin annular region. The authors attributed this phenomenon to a transverse force, which is centrifugal at the channel center and centripetal at the walls. In the absence of inertial and non-Newtonian effects, Saffman~\cite{saffman1956motion} pointed out that a rigid spherical particle in a creeping flow does not experience any sideways force in flows with a unidirectional velocity. Goldsmith \& Mason~\cite{goldsmith1962flow} conducted similar experiments to the one of Segr\'{e} \& Silberberg, but under a creeping flow condition and observed that the cross-stream migration occurs only for non-rigid spherical particles. A follow-up study of the same authors~\cite{karnis1963axial} shows a cross-streamline inward migration of rigid spheres in non-Newtonian fluids. Bretherton showed theoretically that rigid particles with specific extreme shapes can experience a lateral drift across streamlines without inertial and non-Newtonian effects~\cite{bretherton1962motion}. Several subsequent theoretical works, based on the method of matching asymptotic expansions, investigated systems differing from that of  Segr\'{e} \& Silberberg in Reynolds number ($Re$), ratio of the particle radius to the pipe radius ($a/R$), bare fluid flow, or channel geometry.\cite{Saffman65,Ho:Leal74,Vas:Cox76,Asmolov99} Interestingly, Ho \& Leal~\cite{Ho:Leal74} found that the lateral equilibrium position of a small particle ($a/R \ll 1$) in a planar two-dimensional Poiseuille flow is identical to the value measured by Segr\'{e} \& Silberberg along the radial direction of the cylindrical pipe. Furthermore, Asmolov~\cite{Asmolov99} pointed out that the equilibrium position in a shear flow shifts further toward the walls as $Re$ increases. In a follow-up study, Asmolov and co-workers \cite{asmolov2018inertial} provided an expression for the lift force on a point-like particle in a confined shear flow, which is asymptotically valid independently from the lateral particle position provided that $a/R \ll 1$ and $Re \lessapprox 10$. In addition, they found a qualitative agreement with numerical simulations of a finite-size particle using the lattice Boltzmann method.

Matas, Morris \& Guazzelli \cite{matas2004} studied the inertial migration of a dilute suspension of particles in a cylindrical channel for a wide range of $Re$ and $a/R$ experimentally. They observed that in addition to the outer annulus reported by Segr\'{e}-Silberberg, particles accumulate into an inner annulus, which the existing theory did not predict. The inner annulus was seen for $Re > 400$ for $a/R \approx 0.05-0.1$, and for $Re > 1200$ for $a/R \approx 0.02$. Matas observed that the particles migrate gradually from the outer to the inner annulus as Re increases until only observing the inner annulus at $Re$ close to intermittency ($Re\approx1650$ for $a/R\approx0.05$). Shao, Yu \& Sun \cite{shao2008} studied the radial migration of a rigid sphere in a pipe flow for $a/R \approx 0.1-0.15$ and $Re$ up to $2200$. They found two distinct equilibrium positions corresponding to the Segr\'{e}-Silberberg and the inner annuli in agreement with the results of Matas. However, numerical simulations showed a clear transition from one equilibrium position to the other as a function of $Re$, which is in contrast with the experiment of Matas, where both equilibrium positions can coexist at a range of $400 \lessapprox Re \lessapprox 1000$). Morita, Itano \& Sugihara-Seki conducted similar experiments to Matas while considering more extended pipe geometries \cite{morita2017}. They observed that the concentration of particles accumulated in the inner annulus decreases further downstream until most of the particles migrate toward the Segr\'{e}-Silberberg annulus provided the channel is long enough. Recently, Pan, Li \& Glowinski \cite{pan2021numerical} reported a turning point bifurcation with an unstable branch between both annuli (stable branches) that spans over a few tens of $Re$ for a large particle ($a/R=0.15$) and over a few hundreds of $Re$ for a small particle ($a/R=0.084$).

Here, we focus on the role played by the channel length by measuring the channel focusing length (also known as entry length) required to observe radial focusing of a dilute suspension of particles. We simulate a dilute suspension of finite size particles with a volume fraction $\phi$ up to $1\%$, $a/R=0.1$ and $Re$ up to $600$. This choice of parameters is motivated by our interest in studying the existence of the unstable branch discussed in the experiment of Matas \cite{matas2004}.

\section{Numerical scheme}
\label{sec:method}
This section provides a summary of the numerical method and the simulation setup. More details on the simulation method are available in the appendix.

\begin{figure}[ht!]
	\begin{center}
		\includegraphics[width=0.8\linewidth]{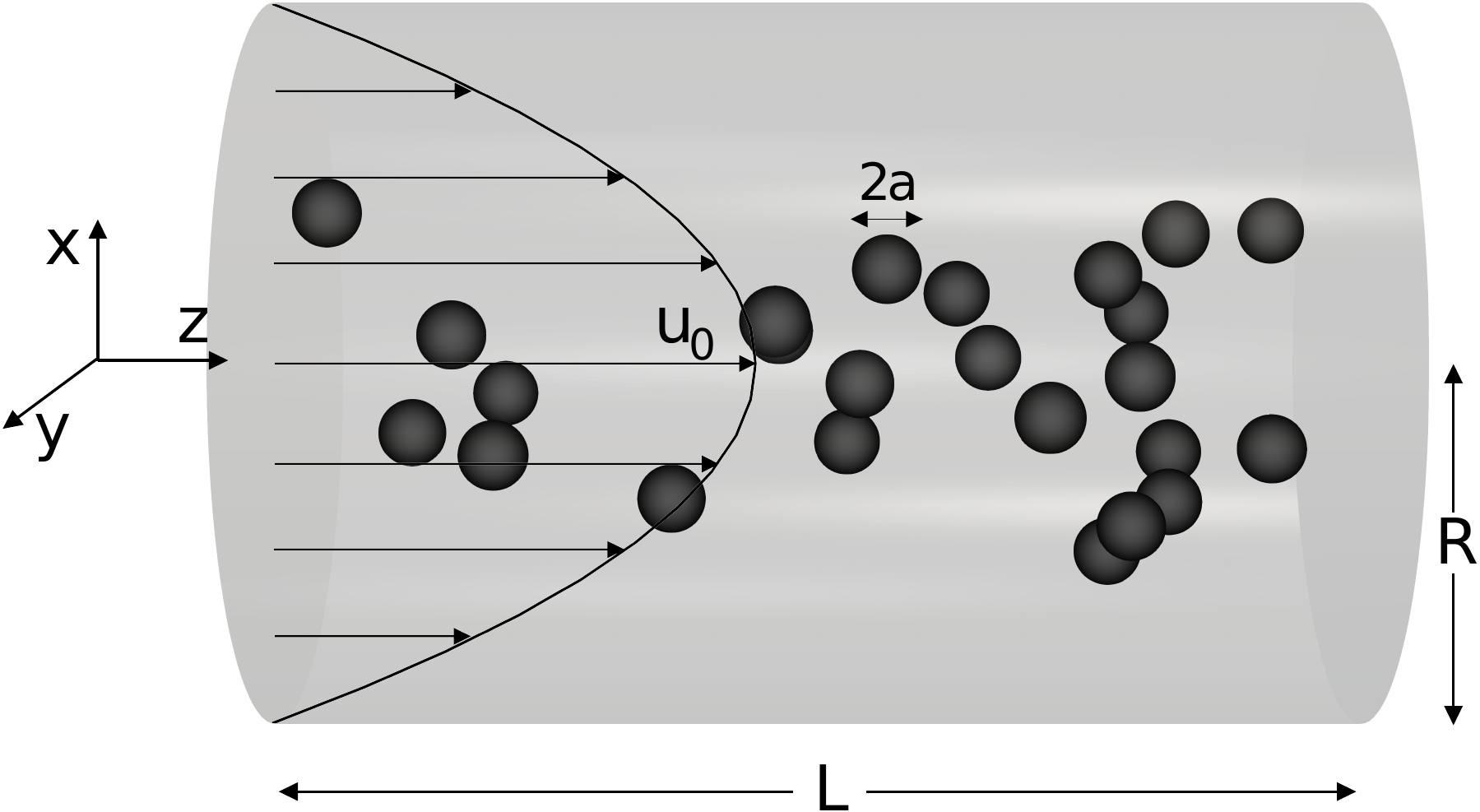}
	\end{center}
	\caption{Schematic of the numerical setup used throughout this work.}
	\label{fig:fig1}
\end{figure}

We consider a suspension of rigid spherical particles of radius ($a$) subject to a parabolic flow in a pipe with a radius ($R$) and a length ($L$). The system is periodic along the $z-$axis, corresponding to the flow direction. The particles are initially homogeneously distributed within the pipe cross-section. The pipe radius $R=10a$ and length $L \approx 3R=30a$. We impose a Poiseuille flow with a centerline velocity ${u}_0$ by applying a body force along the $z$-direction as depicted in Fig\ref{fig:fig1}.
The steady-state radial position of the particles ($r_{s}$) in the cylindrical channel is mainly governed by the channel Reynolds number ($Re=\frac{{u}_0 R}{\nu}$), which quantifies the importance of the inertial forces over the viscous forces ($\nu$ is the kinematic viscosity of the fluid), and the particle-to-channel radius ($a/R$). We set the volume fraction of the suspension to $\phi \le 1\%$ to minimize the effect of particle-particle hydrodynamic interactions.
Each simulation was carried out on 200 CPUs on the high-performance computing clusters available at the Erlangen National High Performance Computing Center (NHR@FAU) or the J\"ulich Supercomputing Center (JSC).

The motion of the fluid is described by the discrete lattice Boltzmann equation in velocity space that reads as
\begin{equation}
  f_i({\mathbf{x}}+\mathbf{c}_i\Delta t, t + \Delta t) - f_i({\mathbf{x}},t) = \Omega_i({\mathbf{x}},t) + F_i({\mathbf{x}},t) \Delta t,
  \label{eq:lbe}
\end{equation}
where $f_i$ is the discrete probability function of finding a particle at position ${\mathbf{x}}$ and time $t$ moving with velocity $\mathbf{c}_i$, $i=1,...,19$ on a 3D Eulerian grid \cite{benzi1992lattice,succi2001,chin2003steering,kruger2017lattice}. Here, $\Omega_i = -\frac{\Delta t}{\tau} [f_i({\mathbf{x}},t)-f_i^{eq}({\mathbf{x}},t)]$ is the standard Bhatnagar-Gross-Krook (BGK) collision operator \cite{bhathnagor1954model}, $\tau$ is the relaxation time, and $f_i^{eq}$ is a truncated expansion of the Maxwell-Boltzmann distribution for the particle velocities in an ideal gas and corresponds to the local equilibrium distribution function. The lattice constant and the discrete time step are denoted by $\Delta x$ and $\Delta t$, respectively. The LBM follows a stream and collide scheme and thus can be divided into two steps: i) the collision step
%in which the particle discrete probability function collide and then relax on the Eulerian grid according to
\begin{equation}
    \tilde{f}_i({\mathbf{x}},t) = f_i({\mathbf{x}},t) -\frac{\Delta t}{\tau} [f_i({\mathbf{x}},t)-f_i^{eq}({\mathbf{x}},t)],
    \label{eq:lbm_collision}
\end{equation}
where $\tilde{f}$ is the post-collision distribution function, and ii) the advection step where the discrete particle probability distributions are streamed from one lattice node to the next one according to their velocities
\begin{equation}
    f_i({\mathbf{x}}+\mathbf{c}_i\Delta t, t + \Delta t) = \tilde{f}_i({\mathbf{x}},t).
    \label{eq:lbm_streaming}
\end{equation}
To generate a parabolic flow, we apply a body force on the fluid along the $z-$axis, incorporated into Eq.~\ref{eq:lbe} through the source term $F_i$.
\begin{equation}
   F_i({\mathbf{x}},t) = \left(1 - \frac{1}{2\tau} \right) \omega_i \left(\frac{\mathbf{c}_i-\mathbf{u}}{c_s^2}+\frac{\mathbf{c}_i \cdot \mathbf{u}}{c_s^4}\mathbf{c}_i\right)\cdot\mathbf{f}({\mathbf{x}},t),
\end{equation}
where $\mathbf{f}({\mathbf{x}},t)$ accounts for the body force, $\mathbf{u}$ is the macroscopic velocity, $c_s = \sqrt{1/3}\Delta x / \Delta t$ is the lattice speed of sound, and $\omega_i$ are the lattice weights which, for the D3Q19 LBM employed here, read as $1/3$, $1/18$ and $1/36$ for $i=1$, $i=2\dots7$, and $i=8\dots19$, respectively. The fluid density ($\rho$) and the macroscopic velocity on each lattice node are calculated from the zeroth and first moments of the discrete probability function $f_i$ such that
\begin{equation}
   \rho = \sum_{i=1}^{19} f_i ({\mathbf{x}},t), \quad \text{and} \hspace{0.25cm}  \mathbf{u} = \sum_{i=1}^{19} f_i ({\mathbf{x}},t)\mathbf{c}_i / \rho.
\end{equation}
The kinematic viscosity of the fluid is related to the relaxation time $\tau$ by
\begin{equation}
    \nu = \rho_0 c_s^2\left(\tau - \frac{\Delta t}{2}\right),
\end{equation}
with $\rho_0$ being the mass density.
By applying a body force ($f_z$) mimicking a pressure gradient along the $z$ direction and a no-slip boundary condition, we obtain a parabolic profile with a maximum velocity ${u}_0=\frac{f_z R^2}{4\rho\nu}$.

We restrict our investigation to a monodisperse suspension of rigid spherical particles with a density ratio between the particle ($\rho_{p}$) and the fluid fixed to unity.
The particle's motion is solved numerically using the discrete element method (DEM), a widely employed numerical scheme in engineering and physics\cite{fleissner2007applications}.
The DEM is based on integrating Newton's equations of motion for a rigid body
\begin{align}
    & M \mathbf{\dot{U}} = \mathbf{F}_{tot} \label{eq:newton_motion}\\
    & \mathbf{I} \dot{\Omega} + \Omega \cdot [\mathbf{I} \cdot \Omega] = \mathbf{T}_{tot},
    \label{eq:euler_dynamic}
\end{align}
where $\mathbf{F}_{tot}$ and $\mathbf{T}_{tot}$ are the net force and torque exerted on the solid particle, $\mathbf{I}$ and $M=\rho_{p} (4/3) \pi R^3$ are the inertia tensor and the mass of the particle, respectively. The contributions to the total force (and torque) stem from the hydrodynamic interactions between the fluid and the particle ($\mathbf{F}_H$), and the lubrication force ($\mathbf{F}_{lub}$). Equations \ref{eq:newton_motion} and \ref{eq:euler_dynamic} are integrated using a leapfrog scheme \cite{allen2017computer}.

\section{Results\protect}
\label{sec:results}
\begin{figure}[ht!]
	\begin{center}
		\includegraphics[width=.4\linewidth]{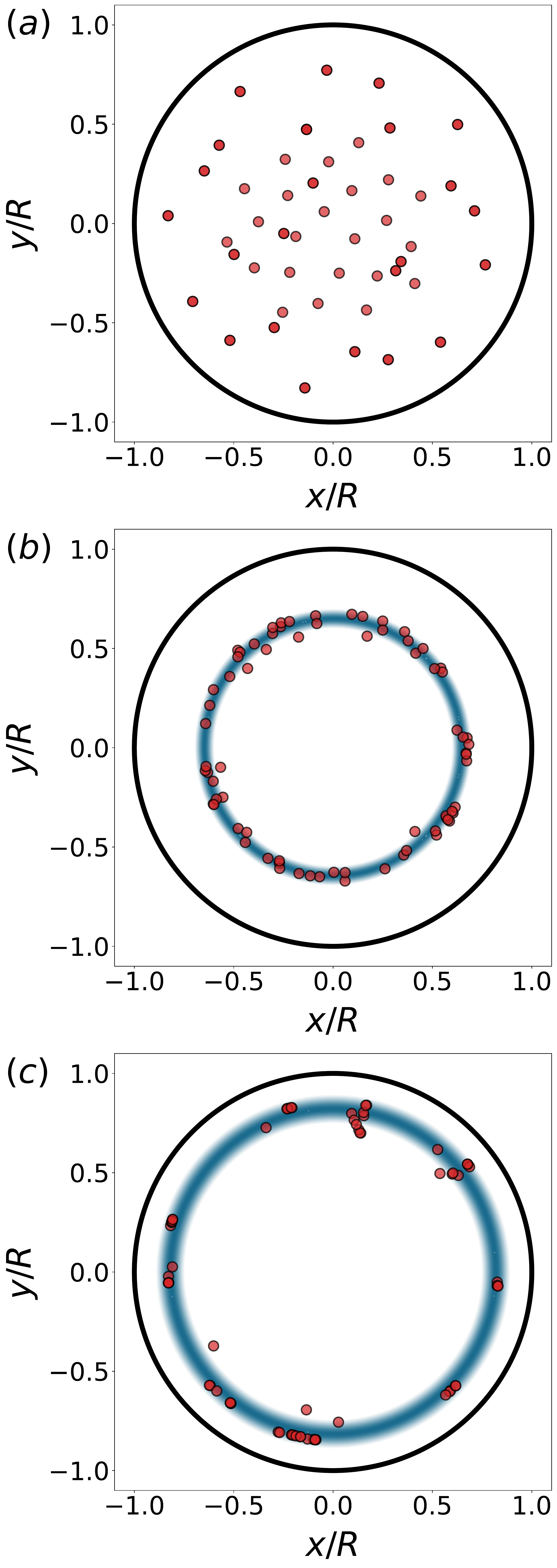}
	\end{center}
	\caption{Particle distribution over the pipe cross-section. (a) Initial distribution of the particles. (b) Steady-state distribution for $Re=38$. (c) Same as (b) but for $Re=600$. The red circles show the location of the mass center of the particles projected over the cross-section.}
	\label{fig:fig2}
\end{figure}

In this section, we present and discuss the role of inertia on the spatial distribution of a dilute suspension of particles for $Re \le 600$ and for a particle-to-pipe radius of $a/R=0.1$. Three volume fractions are considered: $\phi = 0.02\%$ (single particle), $\phi = 0.5\%$ and $\phi=1\%$.
In Fig.~\ref{fig:fig2} we show the radial probability distribution of the particles at $\phi=1\%$ projected over the cross-section, where the radial position of each particle is averaged over time after excluding the transient phase. The transient phase here is the time after which all the particles reach their steady-state radial position.
The particles accumulate into an annular structure under the balance between the shear gradient and the wall lift forces. This phenomenon corresponds to the tubular pinch effect discussed in the experimental work of Segr\'{e} and Silberberg \cite{segre1961,segre1962jfm2}.
The radius of the annulus increases with $Re$, which can be correlated to increasing the shear-gradient lift force on the particles.

\begin{figure}[ht!]
	\begin{center}
		\includegraphics[width=.5\linewidth]{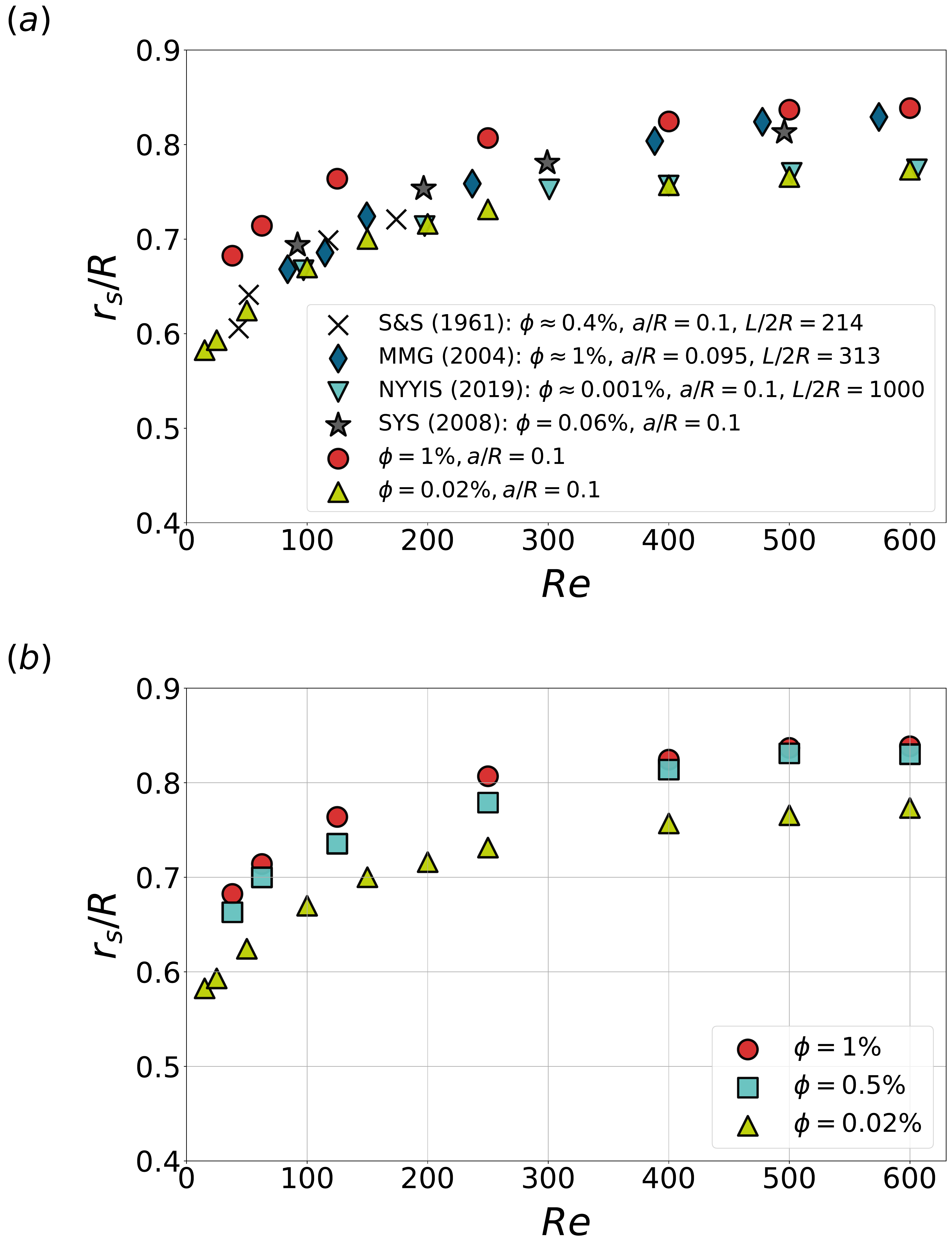}
	\end{center}
	\caption{Steady-state radial position as a function of $Re$. (a) Our numerical results for a single particle and a suspension of particles are depicted with triangles and circles. Experimental results from S\'egre \& Silberberg \cite{segre1961} (S\&S), Matas \textit{et al.} \cite{matas2004} (MMG), and Nakayama \textit{et al.} \cite{sugihara-seki2019} (NYYIS) are indicated by crosses, diamonds, and inverted triangles, respectively. The numerical data from Shao \textit{et al.} \cite{shao2008} (SYS) are represented with stars. (b) Steady-state radial position as function of $Re$ for different volume fractions.
	}
	\label{fig:fig3}
\end{figure}

In the following, we will start by comparing our results for the radial migration with some of the existing works from literature for $a/R=0.1$. Most of the experiments and numerical simulations were performed in the dilute regime, typically up to $\phi = 1\%$, to minimize the effect of the hydrodynamic interaction between particles. In Fig.~\ref{fig:fig3} we report the steady-state radial position as a function of $Re$ and $\phi$.
In our numerical simulations, we compute the steady-state radial position ($r_s/R$) by averaging $r/R$ over time (excluding the transient phase) and over the number of particles. Experimental measures of $r_s/R$ are typically performed at different axial positions ($L/2R$) across the length of the channel. This procedure is equivalent to taking snapshots of the particle distribution at a given axial position with the disadvantage of not having access to the radial position history and, thus, a strict criterion to define the steady-state. We define the steady-state as the radial position of all the particles reaching a plateau. For the case of a single particle (green triangles), the steady-state radial position increases with $Re$, then starts to saturate at $400 \le Re \le 600$, approaching an asymptotic limit of $r_s/R \approx 0.77$ at $Re = 600$. The suspension of particles (red circles) exhibits a similar behavior but with a slightly higher asymptotic limit of $r_s/R \approx 0.83$. We present the effect of the volume fraction in Fig.~\ref{fig:fig3}(b). Although we are in the dilute regime ($\phi \le 1\%$), we observe that $r_s/R$ is typically larger for suspensions ($\phi=0.5\%$ and $\phi=1\%$) when compared to a single particle ($\phi = 0.02\%$) which suggests that we cannot neglect the hydrodynamic interaction between particles in the inertial regime. We also observe a similar behavior by comparing to the experiments of MMG\cite{matas2004} and NYYIS\cite{sugihara-seki2019} at, respectively, $\phi \approx 0.1\%$ and $\phi \approx 0.001\%$.

\begin{figure}[ht!]
	\begin{center}
		\includegraphics[width=.5\linewidth]{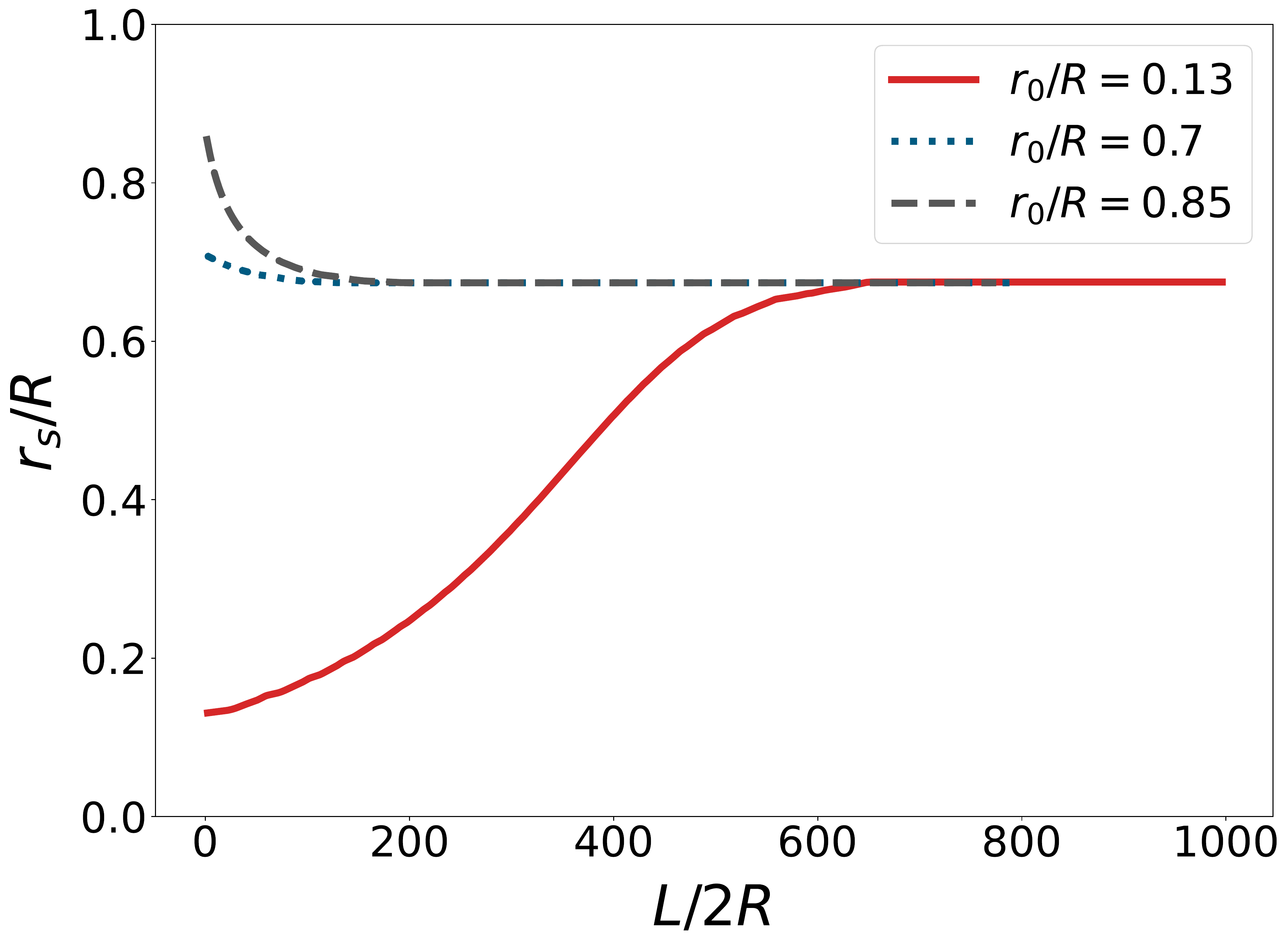}
	\end{center}
	\caption{Radial position history as a function of the axial downstream position for the case of a single particle placed at different initial positions $r_0/R=0.13$, $0.7$ and $0.85$, respectively. The remaining parameters are kept fixed at $Re=50$, $\phi=0.02\%$ and $a/R=0.1$.}
	\label{fig:fig4}
\end{figure}

\begin{figure}[ht!]
	\begin{center}
		\includegraphics[width=.5\linewidth]{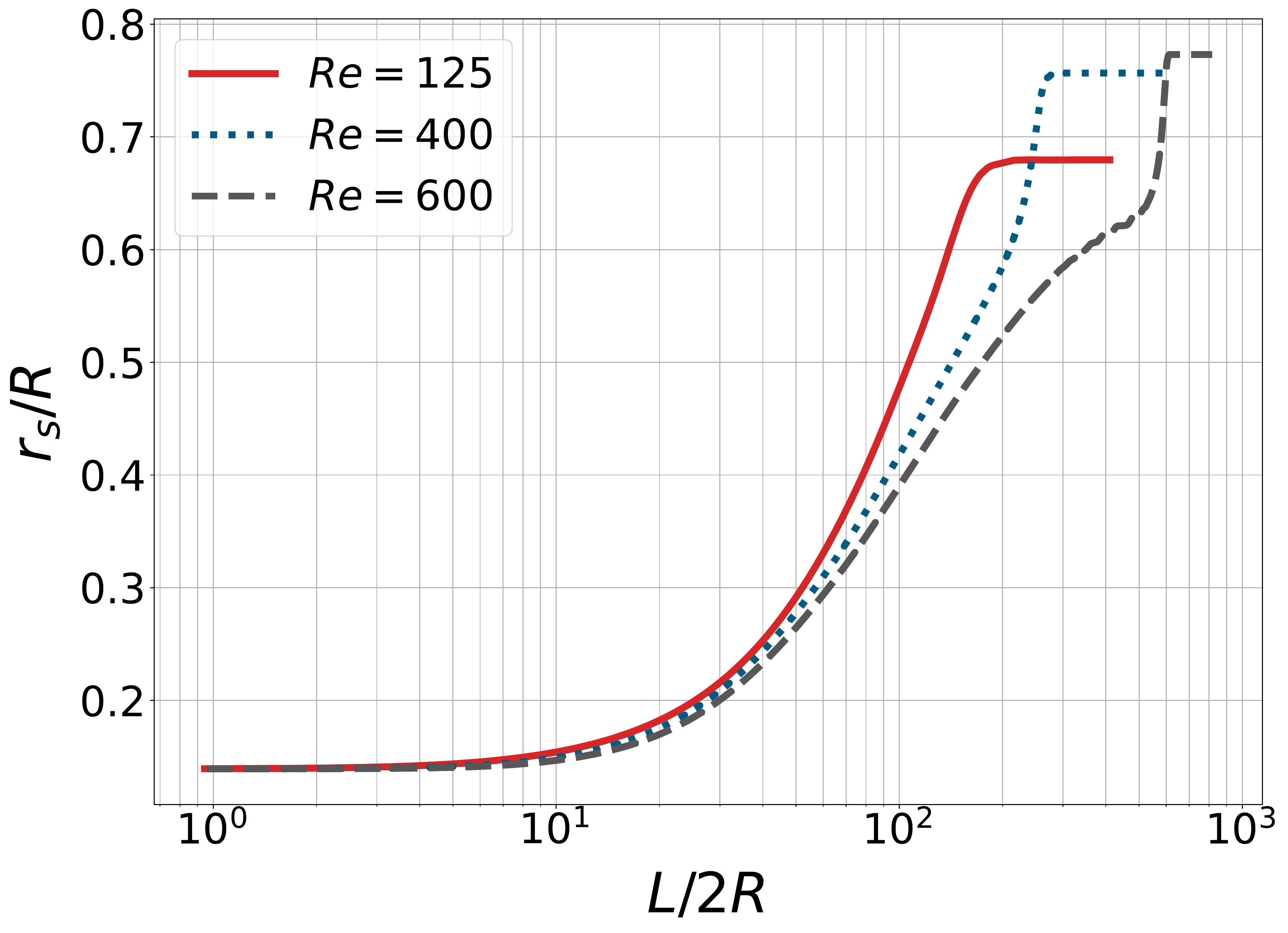}
	\end{center}
	\caption{Radial position history of a single particle ($\phi=0.02\%$) with respect to the axial downstream position for different values of $Re$. Note that the x-axis is represented in log scale.}
	\label{fig:fig5}
\end{figure}

\begin{figure}[ht!]
	\begin{center}
		\includegraphics[width=.5\linewidth]{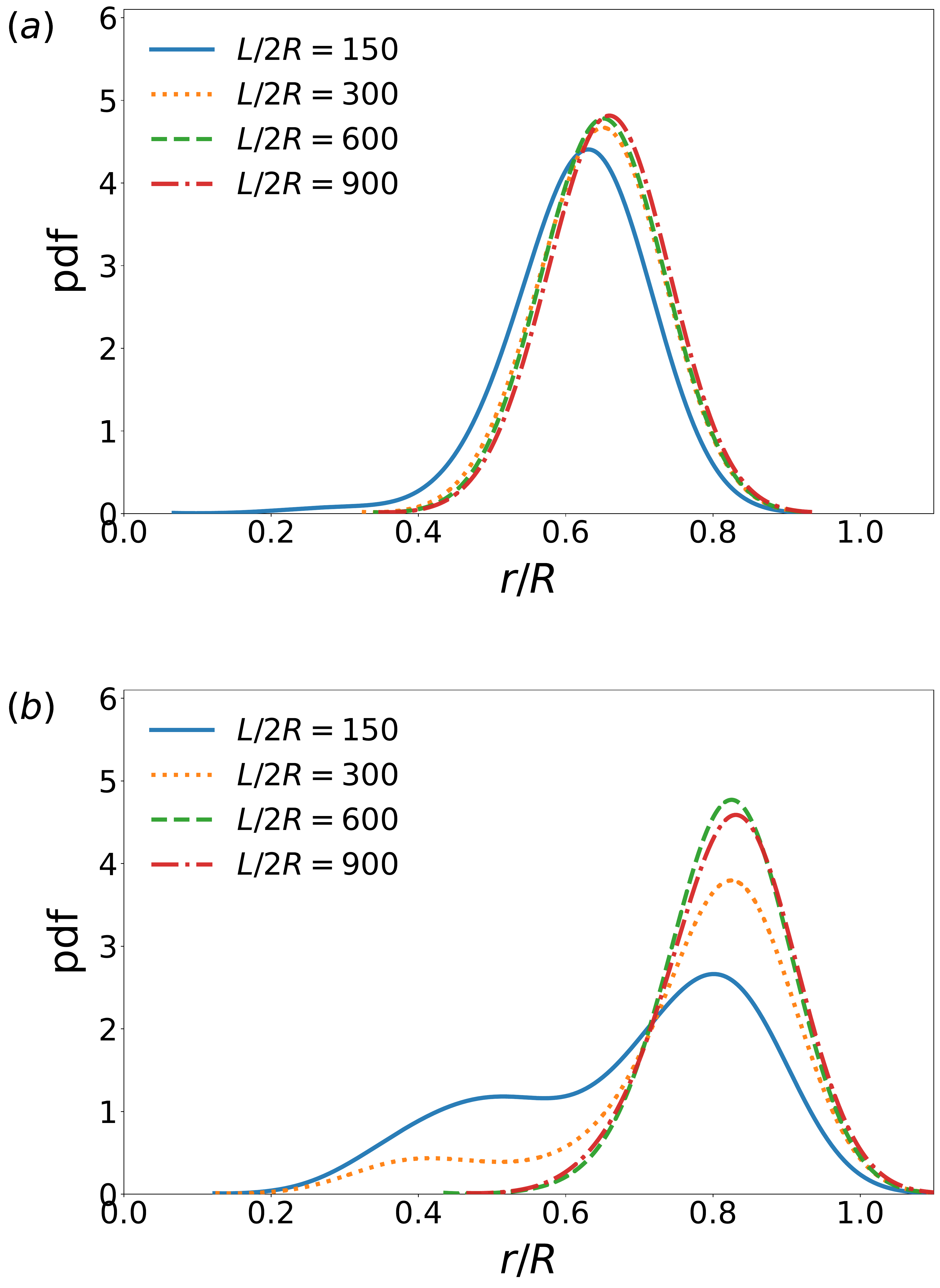}
	\end{center}
	\caption{Probability distribution function of the radial position of the particles at different axial positions $L/2R$ for $Re=38$ (a) and $Re=600$ (b). $\phi$ is fixed to $1\%$}
	\label{fig:fig6}
\end{figure}

\begin{figure}[ht!]
	\begin{center}
		\includegraphics[width=.5\linewidth]{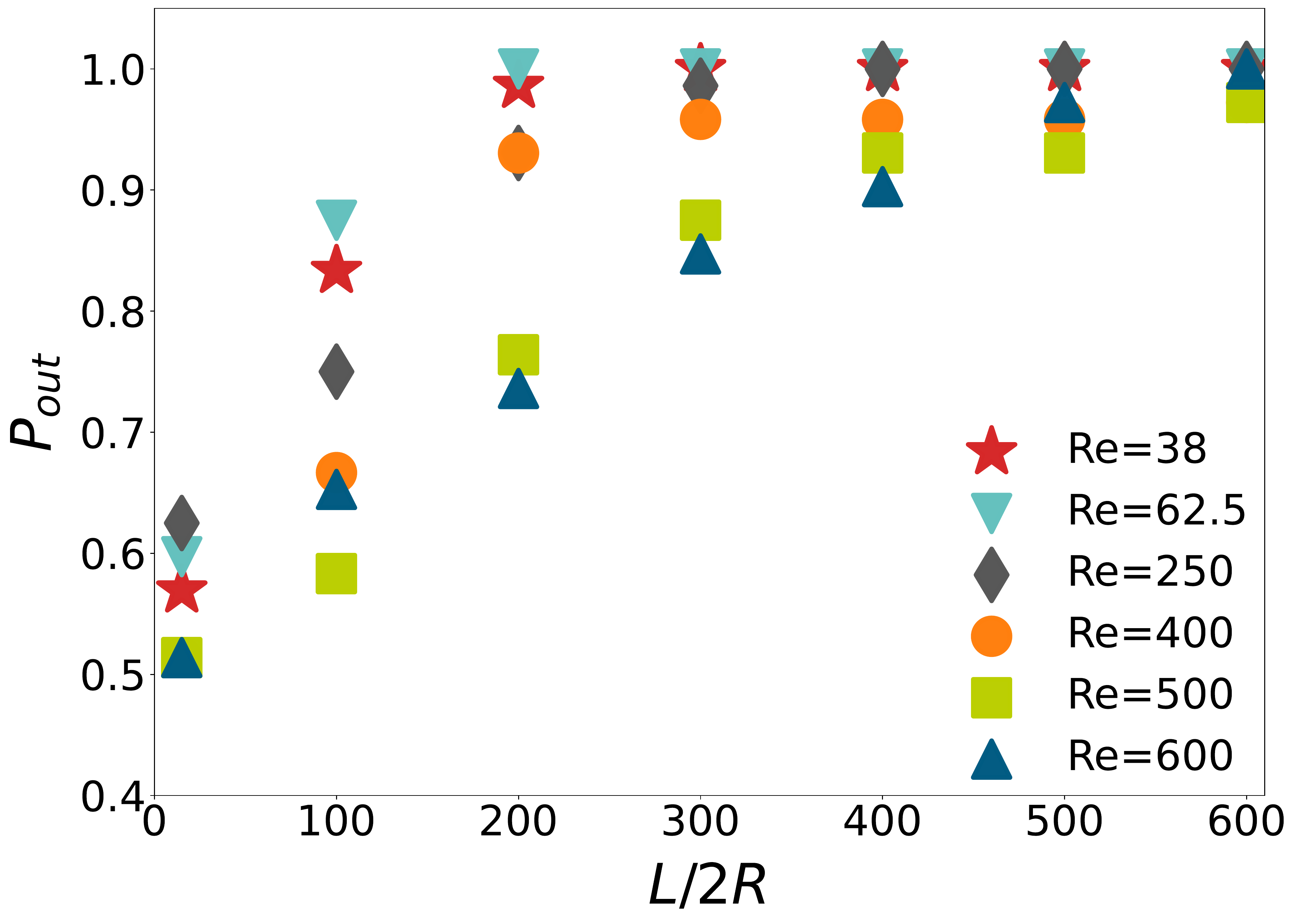}
	\end{center}
	\caption{Probability of finding particles in the outer annulus as a function of the axial position of the cross-section $L/2R$, for different values of $Re$ and for $\phi$ fixed to $1\%$.}
	\label{fig:fig7}
\end{figure}

\begin{figure*}[ht!]
	\begin{center}
		\includegraphics[width=.9\linewidth]{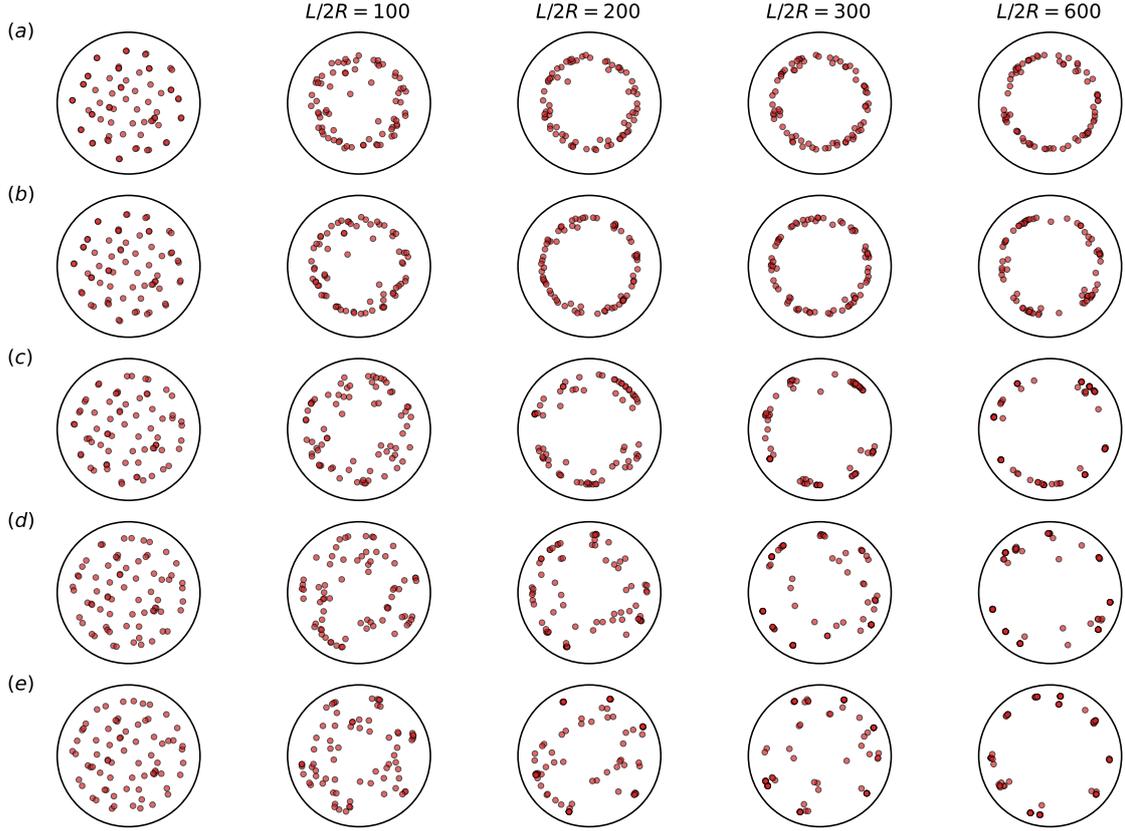}
	\end{center}
	\caption{Particle distribution over the pipe cross-section at different axial positions $L/2R$ and for different values of $Re$: (a) $Re=38$, (b) $Re=62.5$, (c) $Re=250$, (d) $Re=500$, and (e) $Re=600$.}
	\label{fig:fig8}
\end{figure*}

In order to exclude the effect of hydrodynamic interactions between particles, we now consider the case of an isolated particle. Looking at the radial position history in Fig.~\ref{fig:fig4}, we find that the particle always migrates toward the same steady-state radial position independently of its initial position. This behavior is persistent for the parameter space considered in this study and defined by $Re \le 600$ and $a/R=0.1$.
We now turn to the channel focusing length, an important observable that quantifies the channel length required for the particle to reach its radial steady-state position. Matas \textit{et al.} \cite{matas2004} estimated the channel focusing length to be inversely proportional to the channel Reynolds number in straight rectangular geometries.  In a subsequent study, Matas \textit{et al.} \cite{matas2009} observed a similar behavior for pipe geometries, although the decrease of the focusing length with the increase of $Re$ was found to be less pronounced when compared to rectangular geometries. The estimate of the focusing length was based under the assumption that $a/R \ll 1$, and therefore  $Re_p = Re(a/R)^2 \ll 1$. Using the finite element method, Di Carlo~\cite{di2009particle} found for a finite-size particle that the scaling of the lift force depends on the position of the particle in the channel and on the particle size.
To further investigate this phenomenon, we start by considering the case of an isolated particle initially located at the same radial position and measure its radial trajectory while varying $Re$. The results are depicted in Fig.~\ref{fig:fig5} for $Re=125$, $400$, and $600$.
Interestingly, and in contrast to the analytical predictions, we observe an increase of the channel focusing length with $Re$. Indeed using matched asymptotic expansions, Matas~\cite{matas2004} predicted the focusing length to decrease with $Re$, such that $L_{s}/2R = 6\pi A^{-1} Re^{-1} (R/a)^{3}$, where $A$ is the magnitude of the scaled radial force. Our results show that this estimate of the focusing length does not hold for a particle with finite size. It is unclear if the reason of the disagreement is related to the particle Reynolds number $Re_p$ being of the order of $\mathcal{O}(1)$ or if it is due to the curved geometry. But in any case, the experiments are often performed at $Re_p=\mathcal{O}(1)$, where the assumption that the flow field is undisturbed by the particle is questionable. This raises the question on the validity of the choice of the length of the channel in experiments based on the estimate from the result of the matched asymptotic expansion. Now that we have a better understanding of the radial migration and focusing length at the scale of a single particle, we move toward the case of a dilute suspension of particles. To investigate the focusing length, we start by measuring the radial probability distribution of the particles at different downstream axial positions $L/2R$, similar to the protocols used in the experimental studies\cite{matas2004,morita2017,sugihara-seki2019}. For that, we have considered three set of simulations with different initial positions. The number of particles for each simulation is $24$. We have computed the probability distribution for each set of data separately and then for the combined three set of data to improve the statistics. In both cases, we have observed the same behavior that is reported in Fig.~\ref{fig:fig6} for $Re=38$ and $Re=600$. It can be seen that the particles reach the radial steady state already after a focusing length of approximately $L/2R = 300$ at $Re=38$. At Re=600 on the other hand we here still observe the remnants of a bimodal distribution, corresponding to an inner and outer annulus, as initially reported in the experimental work of Matas \cite{matas2004} in pipes. As the particles travel further downstream, the inner peak disappears and the particles accumulate at the outer (Segr\'{e}-Silberberg) annulus, which suggests that the channel focusing length for $Re=600$ and $a/R=0.1$ is beyond $L/2R=300$ and that any measure of the radial steady-state should be done further downstream.
A second interesting observation is that the channel focusing length ($L_s/2R$) for $Re=600$ is larger than for $Re=38$, suggesting that $L_s/2R$ increases with $Re$, similar to the single particle case. This unexpected results was also observed in the experimental work of Nakayama \textit{et al.} \cite{sugihara-seki2019} for $Re$ up to $600$ but not discussed further.
To explain this behaviour, we report in Fig.~\ref{fig:fig7} the probability of finding the particles at the outer ($P_{out}$) annulus at different downstream axial positions. The location of the outer annulus is known from the radial steady-state measurement performed in Fig.\ref{fig:fig3}.
We observe that independently from $Re$, the probability of the particles accumulated around the channel center decreases as $L/2R$ increases until the focusing length is reached, where the particles are fully migrated in the outer annulus. Our simulations clarify that the inner annulus reported by Matas \textit{et al.}~\cite{matas2004}, and in the parameter space investigated here, is a transient configuration which is in agreement with the recent experimental work of Nakayama \textit{et al.}~\cite{sugihara-seki2019} using longer pipes with $L/2R$ up to $1000$. Interestingly, the existence of the inner annulus seems to persist at even further downstream axial positions with the increase of $Re$, which may explain the increase of the focusing length with $Re$ in pipe geometries. This can be seen clearly in Fig.~\ref{fig:fig8}, where we show snapshots of the radial particle distribution over the pipe cross-section at different axial positions and for different $Re$.
We can observe the appearance of the inner annulus at relatively large $Re$ at $L/2R \le 300$ and its full disappearance at a further downstream location, here chosen at $L/2R=600$.

\section{Conclusion\protect}
\label{sec:conclusion}
We have studied numerically the inertial migration of an isolated particle and a dilute suspension of particles ($\phi \le 1\%$) in a pipe flow for $Re$ up to $600$ and $a/R=0.1$. Our simulations shed light on the particle dynamics by continuously monitoring them on their downstream propagation. We found that in the dilute regime, the particles migrate radially and form a structured ring located between the center and channel walls corresponding to the tubular pinch effect reported by Segr\'{e} and Silberberg \cite{segre1961}. The radius of the ring increases with $Re$ and reaches an asymptotic limit at around $Re \ge 600$. Interestingly for the suspension case, the steady-state radial position $r_s$ is further away toward the wall than in the single-particle case. At the same time, we do not observe a significant difference between volume fractions of $\phi=0.5\%$ and $\phi=1\%$. This suggests that the pair hydrodynamic interactions are non-negligible even in the dilute regime. This observation was confirmed when measuring the probability distribution function of the particles at different downstream axial positions, where we observed that the increase of $Re$ leads to the accumulation of some of the particles in a transient inner annulus close to the  centerline while the rest of the particles form an outer annulus located close to the wall. At further downstream positions, the inner annulus disappears, and the particles are all located in the outer annulus. This confirms the experimental results of Nakayama \textit{et al.}~\cite{sugihara-seki2019} on the transient nature of the inner annulus observed by Matas \textit{et al.}~\cite{matas2004} at $L/R=313$. Interestingly, we found that the channel focusing length increases with $Re$, unlike the analytical predictions for a point-like particle in rectangular~\cite{di2009inertial} and pipe~\cite{matas2009} geometries. The experiments on longer pipes\cite{sugihara-seki2019} support our result, although the authors do not explicitly discuss it. Thus, we speculate that the disagreement with the analytical model stems from the particle being sufficiently large to disturb the flow on the one hand and the pipe geometry on the other hand.

\begin{acknowledgments}
This work was supported by the Deutsche Forschungsgemeinschaft (DFG) within the research unit FOR2688 ``Instabilities, Bifurcations and Migration in Pulsatile Flows'' by the grants (HA4382/8-1) and (AV156/1-1), as well as SFB1411, project-ID 416229255.
The authors gratefully acknowledge the scientific support and HPC resources provided by the Erlangen National High Performance Computing Center (NHR@FAU) of the Friedrich-Alexander-Universität Erlangen-Nürnberg (FAU) and the J\"ulich Superomputing Center.
\end{acknowledgments}

\section*{Conflict of Interest}
The authors have no conflicts to disclose.

\section*{Data Availability}
The data that support the findings of this study are available from the authors upon reasonable request.

\appendix

\section{Details about the numerical method}

\subsection{Boundary conditions and solid-fluid coupling}
The hydrodynamic interaction between the solid particles and the fluid is based on the coarse-grained approach developed by Ladd and Aidun~\cite{ladd-verberg2001,aidun1998direct}, where the curved surface of the solid particle is discretized as a zigzag staircase which can be considered as a smooth surface provided the grid mesh is fine enough \cite{ladd1994numerical-1}.
We apply a mid-link bounce-back rule (BB) with the momentum correction when the fluid is advected onto the surface of the particle such that
\begin{equation}
    f_{\bar{i}}({\mathbf{x}},t + \Delta t) = \tilde{f}_i({\mathbf{x}},t) + 2 \omega_i \rho \frac{\mathbf{c}_{\bar{i}} \cdot \mathbf{u}_b}{c_s^2}.
    \label{eq:mid-link-bounce-back}
\end{equation}
Equation \ref{eq:mid-link-bounce-back} implies that a discrete distribution function moving from a fluid node ($\mathbf{x}$) toward a solid node ($\mathbf{x}_s$) will bounce back, half-way from $\mathbf{x}_s$, on the boundary node ($\mathbf{x}_b$) and travel to the opposite direction with a velocity $\mathbf{c}_{\bar{i}} = -\mathbf{c}_{i}$. The last term in the RHS of the equation \ref{eq:mid-link-bounce-back} accounts for a momentum correction applied when a boundary node $\mathbf{x}_b$ is moving with a velocity $\mathbf{u}_b$ defined as
\begin{equation}
    \mathbf{u}_b = \mathbf{U} + {\Omega} \times [\mathbf{x}_b + \frac{1}{2}\mathbf{c}_i \Delta t - \mathbf{X}],
\end{equation}
where $\mathbf{U}$ and ${\Omega}$ are the solid particle translational and angular velocities respectively, and $\mathbf{X}$ is the solid particle center of mass position. In the case of static solid nodes, e.g. the pipe boundaries, the BB rule is reduced to
\begin{equation}
    f_{\bar{i}}({\mathbf{x}},t + \Delta t) = \tilde{f}_i({\mathbf{x}},t).
    \label{eq:static-bounce-back}
\end{equation}
The no-slip boundary condition is satisfied on the solid boundaries (particles and walls) through the BB rule with second-order accuracy, provided that the grid resolution is high\cite{kruger2017lattice}. Fluid nodes without a single lattice velocity directed toward the solid nodes are typically streamed following Eq.~\ref{eq:lbm_streaming}.

\subsection{Particle dynamics}
The hydrodynamic force applied to the solid particle is obtained using the momentum exchange method (MEM). The hydrodynamic force can be calculated as the sum of the BB collisions over the boundary nodes located mid-way between the fluid and solid nodes. As a result, the hydrodynamic force and torque exerted on the solid particle can be written as
\begin{align}
    & \mathbf{F}_{H} = \frac{\Delta x^3}{\Delta t} \sum_{\mathbf{x}_b}^{} \sum_{i}^{}[\tilde{f}_i({\mathbf{x}},t) + f_{\bar{i}}({\mathbf{x}},t + \Delta t)] \mathbf{c}_i \\
    \label{eq:force_hydro}
    & \mathbf{T}_{H} = \frac{\Delta x^3}{\Delta t} \sum_{\mathbf{x}_b}^{} \sum_{i}^{}(\mathbf{x}_b - \mathbf{X}) \times \mathbf{F}_{H} ,
\end{align}
When the solid particle is advected, the solid nodes on the back of the particle (opposite to the motion's direction) are cleared, and their momentum is transferred to the newly created fluid nodes. A similar procedure is applied to the fluid nodes on the front of the particle that is converted into solid nodes, but this time the momentum is transferred from the fluid to the newly created solid nodes. The covering and uncovering of fluid nodes can lead to a violation of global momentum conservation. Thus a correction term needs to be added to the hydrodynamic force and torque exerted on the particle to fulfill the global momentum conservation. We follow here the approach described by Jansen \& Harting in \cite{jansen2011bijels,KHH04} which the expressions of the hydrodynamic force and torque after including the corrections due to the covering and uncovering of fluid nodes during the particle's motion read as
\begin{align}
    & \mathbf{F}_{H} = \frac{\Delta x^3}{\Delta t}\{\sum_{\mathbf{x}_b}^{} \sum_{i}^{}[\tilde{f}_i({\mathbf{x}},t) + f_{\bar{i}}({\mathbf{x}},t + \Delta t)] \mathbf{c}_i -
    \sum_{\mathbf{x}_{f \rightarrow s}}^{} \rho \mathbf{u} +
    \sum_{\mathbf{x}_{s \rightarrow f}}^{} \rho^{\prime} \mathbf{u}_{b}\},
    \label{eq:force_hydro_corrected}
\end{align}
and
\begin{align}
    \mathbf{T}_{H} &= \frac{\Delta x^3}{\Delta t} \{\sum_{\mathbf{x}_{b}}^{} \sum_{i}^{}(\mathbf{x}_b - \mathbf{X}) \times [\tilde{f}_i({\mathbf{x}},t) + f_{\bar{i}}({\mathbf{x}},t + \Delta t)] \mathbf{c}_i \\
    & - \sum_{\mathbf{x}_{f \rightarrow s}}^{} \sum_{i}^{}(\mathbf{x}_{f \rightarrow s} - \mathbf{X}) \times \rho \mathbf{u} \\
    & + \sum_{\mathbf{x}_{s \rightarrow f}}^{} \sum_{i}^{}(\mathbf{x}_{s \rightarrow f} - \mathbf{X}) \times \rho^{\prime} \mathbf{u}_{b} \}.
    \label{eq:torque_hydro_corrected}
\end{align}
Here $\rho^{\prime}$ is the fluid density of the newly created fluid node calculated by averaging the fluid density of the first set of neighboring nodes \cite{aidun1998direct}.
When the surface-to-surface distance between two solid particles is below one lattice space, the flow in the gap can not be resolved anymore. To overcome this issue, we add a short-range lubrication correction force as discussed in Ref.~\cite{ladd-verberg2001,nguyen2002lubrication}. The lubrication force correction exerted on particle $i$ by the particle $j$ is defined as
\begin{equation}
\mathbf{F}_{lub} = \left\{
\begin{array}{rcr}
&\frac{3 \pi a^2 \rho \nu }{2} \bar{\mathbf{d}}_{ij} \cdot \bar{\mathbf{d}}_{ij} [\mathbf{U}_i-\mathbf{U}_j][\frac{1}{||\delta_{ij}||} - \frac{1}{\delta_{c}}] \quad &\text{if} \quad ||\delta_{ij}|| \le \delta_c \\
&\mathbf{0} \quad &\text{if} \quad ||\delta_{ij}|| > \delta_{c}
\end{array}
\right. ,
\label{eq:flub}
\end{equation}
where $a$ is the radius of the particles, $\mathbf{d}_{ij}$ is the mass-to-mass center distance and $\bar{\mathbf{d}}_{ij} = {\mathbf{d}}_{ij}/{d_{ij}}$ its correspondent unit vector. The lubrication force correction is applied when the gap distance between the particles $||\delta_{ij}|| = d_{ij}-2R$ is smaller than the cut off distance $\delta_{c}$ which is fixed in this work to $2/3$ as suggested by Nguyen \& Ladd \cite{nguyen2002lubrication,ladd-verberg2001}.

\section{Radial history of the particles}
\begin{figure}[ht!]
	\begin{center}
		\includegraphics[width=.5\linewidth]{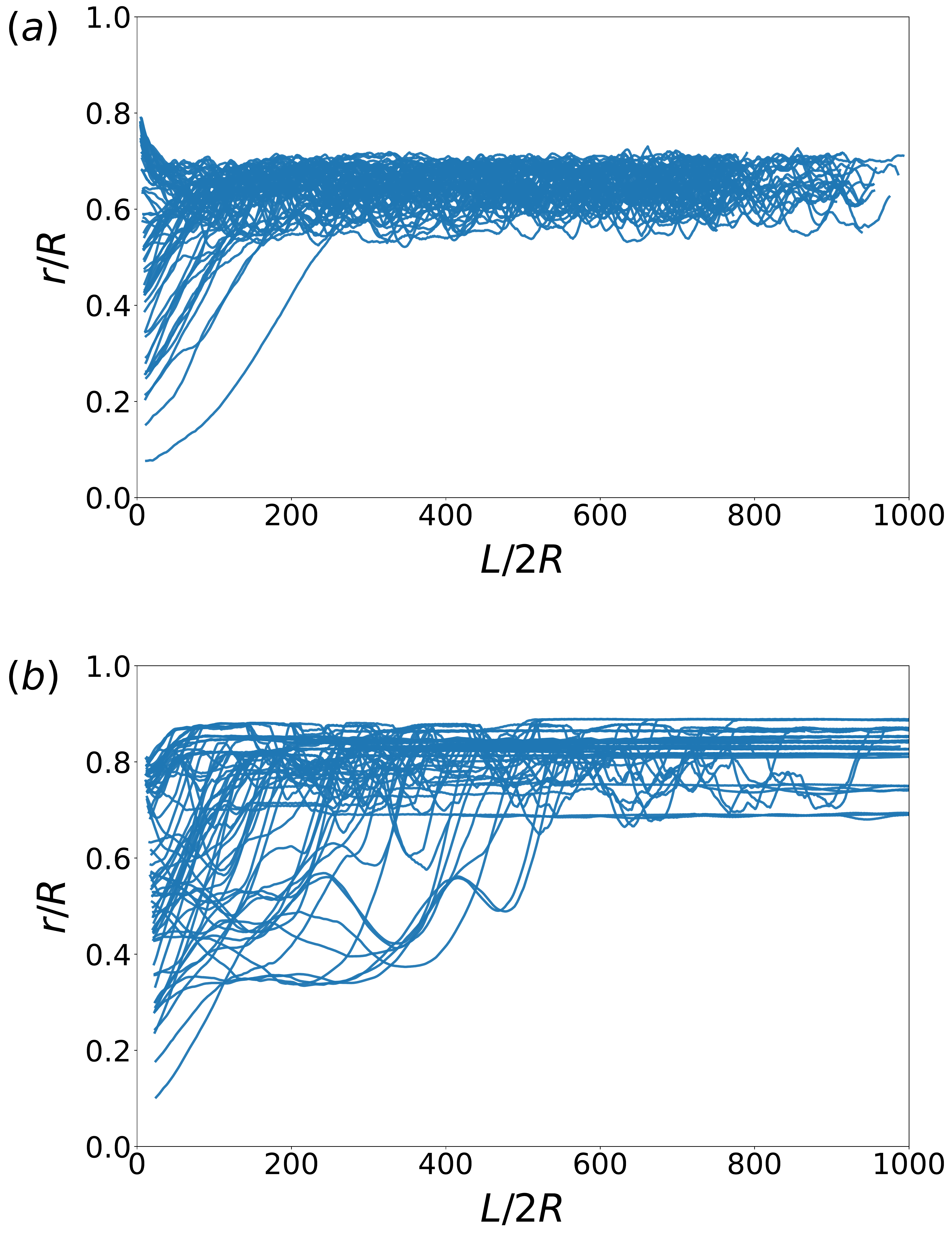}
	\end{center}
	\caption{Radial position history of the particle suspension at $\phi=1\%$ along the flow direction for $Re=38$ (a), and $Re=600$ (b).}
	\label{fig:a2_fig1}
\end{figure}

In Fig. \ref{fig:a2_fig1}, we show the full radial position history of the particles at $\phi=1\%$ for the smallest and largest $Re$ in this study, namely $Re=38$ and $Re=600$.
We observe that the complete radial focusing of the particles at $Re=38$ occurs at a shorter axial distance $L/2R$ as compared to $Re=600$, and that there is no noticeable accumulation of the particles at the inner annulus. Conversely, at $Re=600$, the particles form two rings corresponding to the Segr\'{e} and Silberberg annulus and the inner annulus. The latter disappears completely as the particles travel further downstream along the channel length.

\providecommand{\noopsort}[1]{}\providecommand{\singleletter}[1]{#1}%


\begin{thebibliography}{32}%
\makeatletter
\providecommand \@ifxundefined [1]{%
 \@ifx{#1\undefined}
}%
\providecommand \@ifnum [1]{%
 \ifnum #1\expandafter \@firstoftwo
 \else \expandafter \@secondoftwo
 \fi
}%
\providecommand \@ifx [1]{%
 \ifx #1\expandafter \@firstoftwo
 \else \expandafter \@secondoftwo
 \fi
}%
\providecommand \natexlab [1]{#1}%
\providecommand \enquote  [1]{``#1''}%
\providecommand \bibnamefont  [1]{#1}%
\providecommand \bibfnamefont [1]{#1}%
\providecommand \citenamefont [1]{#1}%
\providecommand \href@noop [0]{\@secondoftwo}%
\providecommand \href [0]{\begingroup \@sanitize@url \@href}%
\providecommand \@href[1]{\@@startlink{#1}\@@href}%
\providecommand \@@href[1]{\endgroup#1\@@endlink}%
\providecommand \@sanitize@url [0]{\catcode `\\12\catcode `\$12\catcode
  `\&12\catcode `\#12\catcode `\^12\catcode `\_12\catcode `\%12\relax}%
\providecommand \@@startlink[1]{}%
\providecommand \@@endlink[0]{}%
\providecommand \url  [0]{\begingroup\@sanitize@url \@url }%
\providecommand \@url [1]{\endgroup\@href {#1}{\urlprefix }}%
\providecommand \urlprefix  [0]{URL }%
\providecommand \Eprint [0]{\href }%
\providecommand \doibase [0]{http://dx.doi.org/}%
\providecommand \selectlanguage [0]{\@gobble}%
\providecommand \bibinfo  [0]{\@secondoftwo}%
\providecommand \bibfield  [0]{\@secondoftwo}%
\providecommand \translation [1]{[#1]}%
\providecommand \BibitemOpen [0]{}%
\providecommand \bibitemStop [0]{}%
\providecommand \bibitemNoStop [0]{.\EOS\space}%
\providecommand \EOS [0]{\spacefactor3000\relax}%
\providecommand \BibitemShut  [1]{\csname bibitem#1\endcsname}%
\let\auto@bib@innerbib\@empty
%</preamble>
\bibitem [{\citenamefont {Segr{\'{e}}}\ and\ \citenamefont
  {Silberberg}(1961)}]{segre1961}%
  \BibitemOpen
  \bibfield  {author} {\bibinfo {author} {\bibfnamefont {G.}~\bibnamefont
  {Segr{\'{e}}}}\ and\ \bibinfo {author} {\bibfnamefont {A.}~\bibnamefont
  {Silberberg}},\ }\href {\doibase 10.1038/189209a0} {\bibfield  {journal}
  {\bibinfo  {journal} {Nature}\ }\textbf {\bibinfo {volume} {189}},\ \bibinfo
  {pages} {209} (\bibinfo {year} {1961})}\BibitemShut {NoStop}%
\bibitem [{\citenamefont {Segre}\ and\ \citenamefont
  {Silberberg}(1962)}]{segre1962jfm2}%
  \BibitemOpen
  \bibfield  {author} {\bibinfo {author} {\bibfnamefont {G.}~\bibnamefont
  {Segre}}\ and\ \bibinfo {author} {\bibfnamefont {A.}~\bibnamefont
  {Silberberg}},\ }\href@noop {} {\bibfield  {journal} {\bibinfo  {journal}
  {Journal of Fluid Mechanics}\ }\textbf {\bibinfo {volume} {14}},\ \bibinfo
  {pages} {136} (\bibinfo {year} {1962})}\BibitemShut {NoStop}%
\bibitem [{\citenamefont {Saffman}(1956)}]{saffman1956motion}%
  \BibitemOpen
  \bibfield  {author} {\bibinfo {author} {\bibfnamefont {P.}~\bibnamefont
  {Saffman}},\ }\href@noop {} {\bibfield  {journal} {\bibinfo  {journal}
  {Journal of Fluid Mechanics}\ }\textbf {\bibinfo {volume} {1}},\ \bibinfo
  {pages} {540} (\bibinfo {year} {1956})}\BibitemShut {NoStop}%
\bibitem [{\citenamefont {Goldsmith}\ and\ \citenamefont
  {Mason}(1962)}]{goldsmith1962flow}%
  \BibitemOpen
  \bibfield  {author} {\bibinfo {author} {\bibfnamefont {H.}~\bibnamefont
  {Goldsmith}}\ and\ \bibinfo {author} {\bibfnamefont {S.}~\bibnamefont
  {Mason}},\ }\href@noop {} {\bibfield  {journal} {\bibinfo  {journal} {Journal
  of Colloid Science}\ }\textbf {\bibinfo {volume} {17}},\ \bibinfo {pages}
  {448} (\bibinfo {year} {1962})}\BibitemShut {NoStop}%
\bibitem [{\citenamefont {Karnis}\ \emph {et~al.}(1963)\citenamefont {Karnis},
  \citenamefont {Goldsmith},\ and\ \citenamefont {Mason}}]{karnis1963axial}%
  \BibitemOpen
  \bibfield  {author} {\bibinfo {author} {\bibfnamefont {A.}~\bibnamefont
  {Karnis}}, \bibinfo {author} {\bibfnamefont {H.}~\bibnamefont {Goldsmith}}, \
  and\ \bibinfo {author} {\bibfnamefont {S.}~\bibnamefont {Mason}},\
  }\href@noop {} {\bibfield  {journal} {\bibinfo  {journal} {Nature}\ }\textbf
  {\bibinfo {volume} {200}},\ \bibinfo {pages} {159} (\bibinfo {year}
  {1963})}\BibitemShut {NoStop}%
\bibitem [{\citenamefont {Bretherton}(1962)}]{bretherton1962motion}%
  \BibitemOpen
  \bibfield  {author} {\bibinfo {author} {\bibfnamefont {F.~P.}\ \bibnamefont
  {Bretherton}},\ }\href@noop {} {\bibfield  {journal} {\bibinfo  {journal}
  {Journal of Fluid Mechanics}\ }\textbf {\bibinfo {volume} {14}},\ \bibinfo
  {pages} {284} (\bibinfo {year} {1962})}\BibitemShut {NoStop}%
\bibitem [{\citenamefont {Saffman}(1965)}]{Saffman65}%
  \BibitemOpen
  \bibfield  {author} {\bibinfo {author} {\bibfnamefont {P.~G.}\ \bibnamefont
  {Saffman}},\ }\href {\doibase 10.1017/S0022112065000824} {\bibfield
  {journal} {\bibinfo  {journal} {Journal of Fluid Mechanics}\ }\textbf
  {\bibinfo {volume} {22}},\ \bibinfo {pages} {385} (\bibinfo {year}
  {1965})}\BibitemShut {NoStop}%
\bibitem [{\citenamefont {Ho}\ and\ \citenamefont {Leal}(1974)}]{Ho:Leal74}%
  \BibitemOpen
  \bibfield  {author} {\bibinfo {author} {\bibfnamefont {B.}~\bibnamefont
  {Ho}}\ and\ \bibinfo {author} {\bibfnamefont {L.}~\bibnamefont {Leal}},\
  }\href {\doibase 10.1017/S0022112074001431} {\bibfield  {journal} {\bibinfo
  {journal} {Journal of Fluid Mechanics}\ }\textbf {\bibinfo {volume} {65}},\
  \bibinfo {pages} {365} (\bibinfo {year} {1974})}\BibitemShut {NoStop}%
\bibitem [{\citenamefont {Vasseur}\ and\ \citenamefont
  {Cox}(1976)}]{Vas:Cox76}%
  \BibitemOpen
  \bibfield  {author} {\bibinfo {author} {\bibfnamefont {P.}~\bibnamefont
  {Vasseur}}\ and\ \bibinfo {author} {\bibfnamefont {R.~G.}\ \bibnamefont
  {Cox}},\ }\href {\doibase 10.1017/S0022112076002498} {\bibfield  {journal}
  {\bibinfo  {journal} {Journal of Fluid Mechanics}\ }\textbf {\bibinfo
  {volume} {78}},\ \bibinfo {pages} {385} (\bibinfo {year} {1976})}\BibitemShut
  {NoStop}%
\bibitem [{\citenamefont {Asmolov}(1999)}]{Asmolov99}%
  \BibitemOpen
  \bibfield  {author} {\bibinfo {author} {\bibfnamefont {E.~S.}\ \bibnamefont
  {Asmolov}},\ }\href {\doibase 10.1017/S0022112098003474} {\bibfield
  {journal} {\bibinfo  {journal} {Journal of Fluid Mechanics}\ }\textbf
  {\bibinfo {volume} {381}},\ \bibinfo {pages} {63} (\bibinfo {year}
  {1999})}\BibitemShut {NoStop}%
\bibitem [{\citenamefont {Asmolov}\ \emph {et~al.}(2018)\citenamefont
  {Asmolov}, \citenamefont {Dubov}, \citenamefont {Nizkaya}, \citenamefont
  {Harting},\ and\ \citenamefont {Vinogradova}}]{asmolov2018inertial}%
  \BibitemOpen
  \bibfield  {author} {\bibinfo {author} {\bibfnamefont {E.~S.}\ \bibnamefont
  {Asmolov}}, \bibinfo {author} {\bibfnamefont {A.~L.}\ \bibnamefont {Dubov}},
  \bibinfo {author} {\bibfnamefont {T.~V.}\ \bibnamefont {Nizkaya}}, \bibinfo
  {author} {\bibfnamefont {J.}~\bibnamefont {Harting}}, \ and\ \bibinfo
  {author} {\bibfnamefont {O.~I.}\ \bibnamefont {Vinogradova}},\ }\href
  {\doibase 10.1017/{Journal of Fluid Mechanics}.2018.95} {\bibfield  {journal}
  {\bibinfo  {journal} {Journal of Fluid Mechanics}\ }\textbf {\bibinfo
  {volume} {840}},\ \bibinfo {pages} {613} (\bibinfo {year}
  {2018})}\BibitemShut {NoStop}%
\bibitem [{\citenamefont {Matas}\ \emph {et~al.}(2004)\citenamefont {Matas},
  \citenamefont {Morris},\ and\ \citenamefont {Guazzelli}}]{matas2004}%
  \BibitemOpen
  \bibfield  {author} {\bibinfo {author} {\bibfnamefont {J.-P.}\ \bibnamefont
  {Matas}}, \bibinfo {author} {\bibfnamefont {J.~F.}\ \bibnamefont {Morris}}, \
  and\ \bibinfo {author} {\bibfnamefont {{\'E}.}~\bibnamefont {Guazzelli}},\
  }\href@noop {} {\bibfield  {journal} {\bibinfo  {journal} {Journal of Fluid
  Mechanics}\ }\textbf {\bibinfo {volume} {515}},\ \bibinfo {pages} {171}
  (\bibinfo {year} {2004})}\BibitemShut {NoStop}%
\bibitem [{\citenamefont {Shao}\ \emph {et~al.}(2008)\citenamefont {Shao},
  \citenamefont {Yu},\ and\ \citenamefont {Sun}}]{shao2008}%
  \BibitemOpen
  \bibfield  {author} {\bibinfo {author} {\bibfnamefont {X.}~\bibnamefont
  {Shao}}, \bibinfo {author} {\bibfnamefont {Z.}~\bibnamefont {Yu}}, \ and\
  \bibinfo {author} {\bibfnamefont {B.}~\bibnamefont {Sun}},\ }\href@noop {}
  {\bibfield  {journal} {\bibinfo  {journal} {Physics of Fluids}\ }\textbf
  {\bibinfo {volume} {20}},\ \bibinfo {pages} {103307} (\bibinfo {year}
  {2008})}\BibitemShut {NoStop}%
\bibitem [{\citenamefont {Morita}\ \emph {et~al.}(2017)\citenamefont {Morita},
  \citenamefont {Itano},\ and\ \citenamefont {Sugihara-Seki}}]{morita2017}%
  \BibitemOpen
  \bibfield  {author} {\bibinfo {author} {\bibfnamefont {Y.}~\bibnamefont
  {Morita}}, \bibinfo {author} {\bibfnamefont {T.}~\bibnamefont {Itano}}, \
  and\ \bibinfo {author} {\bibfnamefont {M.}~\bibnamefont {Sugihara-Seki}},\
  }\href@noop {} {\bibfield  {journal} {\bibinfo  {journal} {Journal of Fluid
  Mechanics}\ }\textbf {\bibinfo {volume} {813}},\ \bibinfo {pages} {750}
  (\bibinfo {year} {2017})}\BibitemShut {NoStop}%
\bibitem [{\citenamefont {Pan}\ \emph {et~al.}(2021)\citenamefont {Pan},
  \citenamefont {Li},\ and\ \citenamefont {Glowinski}}]{pan2021numerical}%
  \BibitemOpen
  \bibfield  {author} {\bibinfo {author} {\bibfnamefont {T.-W.}\ \bibnamefont
  {Pan}}, \bibinfo {author} {\bibfnamefont {A.}~\bibnamefont {Li}}, \ and\
  \bibinfo {author} {\bibfnamefont {R.}~\bibnamefont {Glowinski}},\ }\href@noop
  {} {\bibfield  {journal} {\bibinfo  {journal} {Physics of Fluids}\ }\textbf
  {\bibinfo {volume} {33}},\ \bibinfo {pages} {033301} (\bibinfo {year}
  {2021})}\BibitemShut {NoStop}%
\bibitem [{\citenamefont {Benzi}\ \emph {et~al.}(1992)\citenamefont {Benzi},
  \citenamefont {Succi},\ and\ \citenamefont {Vergassola}}]{benzi1992lattice}%
  \BibitemOpen
  \bibfield  {author} {\bibinfo {author} {\bibfnamefont {R.}~\bibnamefont
  {Benzi}}, \bibinfo {author} {\bibfnamefont {S.}~\bibnamefont {Succi}}, \ and\
  \bibinfo {author} {\bibfnamefont {M.}~\bibnamefont {Vergassola}},\
  }\href@noop {} {\bibfield  {journal} {\bibinfo  {journal} {Physics Reports}\
  }\textbf {\bibinfo {volume} {222}},\ \bibinfo {pages} {145} (\bibinfo {year}
  {1992})}\BibitemShut {NoStop}%
\bibitem [{\citenamefont {Succi}(2001)}]{succi2001}%
  \BibitemOpen
  \bibfield  {author} {\bibinfo {author} {\bibfnamefont {S.}~\bibnamefont
  {Succi}},\ }\href@noop {} {\emph {\bibinfo {title} {The lattice {B}oltzmann
  equation for fluid dynamics and beyond}}}\ (\bibinfo  {publisher} {Oxford
  University Press},\ \bibinfo {year} {2001})\BibitemShut {NoStop}%
\bibitem [{\citenamefont {Chin}\ \emph {et~al.}(2003)\citenamefont {Chin},
  \citenamefont {Harting}, \citenamefont {Jha}, \citenamefont {Coveney},
  \citenamefont {Porter},\ and\ \citenamefont {Pickles}}]{chin2003steering}%
  \BibitemOpen
  \bibfield  {author} {\bibinfo {author} {\bibfnamefont {J.}~\bibnamefont
  {Chin}}, \bibinfo {author} {\bibfnamefont {J.}~\bibnamefont {Harting}},
  \bibinfo {author} {\bibfnamefont {S.}~\bibnamefont {Jha}}, \bibinfo {author}
  {\bibfnamefont {P.~V.}\ \bibnamefont {Coveney}}, \bibinfo {author}
  {\bibfnamefont {A.~R.}\ \bibnamefont {Porter}}, \ and\ \bibinfo {author}
  {\bibfnamefont {S.~M.}\ \bibnamefont {Pickles}},\ }\href@noop {} {\bibfield
  {journal} {\bibinfo  {journal} {Contemporary Physics}\ }\textbf {\bibinfo
  {volume} {44}},\ \bibinfo {pages} {417} (\bibinfo {year} {2003})}\BibitemShut
  {NoStop}%
\bibitem [{\citenamefont {Kr{\"u}ger}\ \emph {et~al.}(2017)\citenamefont
  {Kr{\"u}ger}, \citenamefont {Kusumaatmaja}, \citenamefont {Kuzmin},
  \citenamefont {Shardt}, \citenamefont {Silva},\ and\ \citenamefont
  {Viggen}}]{kruger2017lattice}%
  \BibitemOpen
  \bibfield  {author} {\bibinfo {author} {\bibfnamefont {T.}~\bibnamefont
  {Kr{\"u}ger}}, \bibinfo {author} {\bibfnamefont {H.}~\bibnamefont
  {Kusumaatmaja}}, \bibinfo {author} {\bibfnamefont {A.}~\bibnamefont
  {Kuzmin}}, \bibinfo {author} {\bibfnamefont {O.}~\bibnamefont {Shardt}},
  \bibinfo {author} {\bibfnamefont {G.}~\bibnamefont {Silva}}, \ and\ \bibinfo
  {author} {\bibfnamefont {E.~M.}\ \bibnamefont {Viggen}},\ }\href@noop {}
  {\bibfield  {journal} {\bibinfo  {journal} {Springer International
  Publishing}\ }\textbf {\bibinfo {volume} {10}},\ \bibinfo {pages} {4}
  (\bibinfo {year} {2017})}\BibitemShut {NoStop}%
\bibitem [{\citenamefont {Bhatnagar}\ \emph {et~al.}(1954)\citenamefont
  {Bhatnagar}, \citenamefont {Gross},\ and\ \citenamefont
  {Krook}}]{bhathnagor1954model}%
  \BibitemOpen
  \bibfield  {author} {\bibinfo {author} {\bibfnamefont {P.~L.}\ \bibnamefont
  {Bhatnagar}}, \bibinfo {author} {\bibfnamefont {E.~P.}\ \bibnamefont
  {Gross}}, \ and\ \bibinfo {author} {\bibfnamefont {M.}~\bibnamefont
  {Krook}},\ }\href {\doibase 10.1103/PhysRev.94.511} {\bibfield  {journal}
  {\bibinfo  {journal} {Physical Review}\ }\textbf {\bibinfo {volume} {94}},\
  \bibinfo {pages} {511} (\bibinfo {year} {1954})}\BibitemShut {NoStop}%
\bibitem [{\citenamefont {Fleissner}\ \emph {et~al.}(2007)\citenamefont
  {Fleissner}, \citenamefont {Gaugele},\ and\ \citenamefont
  {Eberhard}}]{fleissner2007applications}%
  \BibitemOpen
  \bibfield  {author} {\bibinfo {author} {\bibfnamefont {F.}~\bibnamefont
  {Fleissner}}, \bibinfo {author} {\bibfnamefont {T.}~\bibnamefont {Gaugele}},
  \ and\ \bibinfo {author} {\bibfnamefont {P.}~\bibnamefont {Eberhard}},\
  }\href@noop {} {\bibfield  {journal} {\bibinfo  {journal} {Multibody system
  dynamics}\ }\textbf {\bibinfo {volume} {18}},\ \bibinfo {pages} {81}
  (\bibinfo {year} {2007})}\BibitemShut {NoStop}%
\bibitem [{\citenamefont {Allen}\ and\ \citenamefont
  {Tildesley}(2017)}]{allen2017computer}%
  \BibitemOpen
  \bibfield  {author} {\bibinfo {author} {\bibfnamefont {M.~P.}\ \bibnamefont
  {Allen}}\ and\ \bibinfo {author} {\bibfnamefont {D.~J.}\ \bibnamefont
  {Tildesley}},\ }\href@noop {} {\emph {\bibinfo {title} {Computer simulation
  of liquids}}}\ (\bibinfo  {publisher} {Oxford university press},\ \bibinfo
  {year} {2017})\BibitemShut {NoStop}%
\bibitem [{\citenamefont {Nakayama}\ \emph {et~al.}(2019)\citenamefont
  {Nakayama}, \citenamefont {Yamashita}, \citenamefont {Yabu}, \citenamefont
  {Itano},\ and\ \citenamefont {Sugihara-Seki}}]{sugihara-seki2019}%
  \BibitemOpen
  \bibfield  {author} {\bibinfo {author} {\bibfnamefont {S.}~\bibnamefont
  {Nakayama}}, \bibinfo {author} {\bibfnamefont {H.}~\bibnamefont {Yamashita}},
  \bibinfo {author} {\bibfnamefont {T.}~\bibnamefont {Yabu}}, \bibinfo {author}
  {\bibfnamefont {T.}~\bibnamefont {Itano}}, \ and\ \bibinfo {author}
  {\bibfnamefont {M.}~\bibnamefont {Sugihara-Seki}},\ }\href@noop {} {\bibfield
   {journal} {\bibinfo  {journal} {Journal of Fluid Mechanics}\ }\textbf
  {\bibinfo {volume} {871}},\ \bibinfo {pages} {952} (\bibinfo {year}
  {2019})}\BibitemShut {NoStop}%
\bibitem [{\citenamefont {Matas}\ \emph {et~al.}(2009)\citenamefont {Matas},
  \citenamefont {Morris},\ and\ \citenamefont {Guazzelli}}]{matas2009}%
  \BibitemOpen
  \bibfield  {author} {\bibinfo {author} {\bibfnamefont {J.-P.}\ \bibnamefont
  {Matas}}, \bibinfo {author} {\bibfnamefont {J.~F.}\ \bibnamefont {Morris}}, \
  and\ \bibinfo {author} {\bibfnamefont {E.}~\bibnamefont {Guazzelli}},\ }\href
  {\doibase 10.1017/S0022112008004977} {\bibfield  {journal} {\bibinfo
  {journal} {Journal of Fluid Mechanics}\ }\textbf {\bibinfo {volume} {621}},\
  \bibinfo {pages} {59} (\bibinfo {year} {2009})}\BibitemShut {NoStop}%
\bibitem [{\citenamefont {Di~Carlo}\ \emph {et~al.}(2009)\citenamefont
  {Di~Carlo}, \citenamefont {Edd}, \citenamefont {Humphry}, \citenamefont
  {Stone},\ and\ \citenamefont {Toner}}]{di2009particle}%
  \BibitemOpen
  \bibfield  {author} {\bibinfo {author} {\bibfnamefont {D.}~\bibnamefont
  {Di~Carlo}}, \bibinfo {author} {\bibfnamefont {J.~F.}\ \bibnamefont {Edd}},
  \bibinfo {author} {\bibfnamefont {K.~J.}\ \bibnamefont {Humphry}}, \bibinfo
  {author} {\bibfnamefont {H.~A.}\ \bibnamefont {Stone}}, \ and\ \bibinfo
  {author} {\bibfnamefont {M.}~\bibnamefont {Toner}},\ }\href@noop {}
  {\bibfield  {journal} {\bibinfo  {journal} {Physical Review Letters}\
  }\textbf {\bibinfo {volume} {102}},\ \bibinfo {pages} {094503} (\bibinfo
  {year} {2009})}\BibitemShut {NoStop}%
\bibitem [{\citenamefont {Di~Carlo}(2009)}]{di2009inertial}%
  \BibitemOpen
  \bibfield  {author} {\bibinfo {author} {\bibfnamefont {D.}~\bibnamefont
  {Di~Carlo}},\ }\href@noop {} {\bibfield  {journal} {\bibinfo  {journal} {Lab
  on a Chip}\ }\textbf {\bibinfo {volume} {9}},\ \bibinfo {pages} {3038}
  (\bibinfo {year} {2009})}\BibitemShut {NoStop}%
\bibitem [{\citenamefont {Ladd}\ and\ \citenamefont
  {Verberg}(2001)}]{ladd-verberg2001}%
  \BibitemOpen
  \bibfield  {author} {\bibinfo {author} {\bibfnamefont {A.}~\bibnamefont
  {Ladd}}\ and\ \bibinfo {author} {\bibfnamefont {R.}~\bibnamefont {Verberg}},\
  }\href@noop {} {\bibfield  {journal} {\bibinfo  {journal} {Journal of
  Statistical Physics}\ }\textbf {\bibinfo {volume} {104}},\ \bibinfo {pages}
  {1191} (\bibinfo {year} {2001})}\BibitemShut {NoStop}%
\bibitem [{\citenamefont {Aidun}\ \emph {et~al.}(1998)\citenamefont {Aidun},
  \citenamefont {Lu},\ and\ \citenamefont {Ding}}]{aidun1998direct}%
  \BibitemOpen
  \bibfield  {author} {\bibinfo {author} {\bibfnamefont {C.~K.}\ \bibnamefont
  {Aidun}}, \bibinfo {author} {\bibfnamefont {Y.}~\bibnamefont {Lu}}, \ and\
  \bibinfo {author} {\bibfnamefont {E.-J.}\ \bibnamefont {Ding}},\ }\href@noop
  {} {\bibfield  {journal} {\bibinfo  {journal} {Journal of Fluid Mechanics}\
  }\textbf {\bibinfo {volume} {373}},\ \bibinfo {pages} {287} (\bibinfo {year}
  {1998})}\BibitemShut {NoStop}%
\bibitem [{\citenamefont {Ladd}(1994)}]{ladd1994numerical-1}%
  \BibitemOpen
  \bibfield  {author} {\bibinfo {author} {\bibfnamefont {A.~J.}\ \bibnamefont
  {Ladd}},\ }\href@noop {} {\bibfield  {journal} {\bibinfo  {journal} {Journal
  of Fluid Mechanics}\ }\textbf {\bibinfo {volume} {271}},\ \bibinfo {pages}
  {285} (\bibinfo {year} {1994})}\BibitemShut {NoStop}%
\bibitem [{\citenamefont {Jansen}\ and\ \citenamefont
  {Harting}(2011)}]{jansen2011bijels}%
  \BibitemOpen
  \bibfield  {author} {\bibinfo {author} {\bibfnamefont {F.}~\bibnamefont
  {Jansen}}\ and\ \bibinfo {author} {\bibfnamefont {J.}~\bibnamefont
  {Harting}},\ }\href@noop {} {\bibfield  {journal} {\bibinfo  {journal}
  {Physical Review E}\ }\textbf {\bibinfo {volume} {83}},\ \bibinfo {pages}
  {046707} (\bibinfo {year} {2011})}\BibitemShut {NoStop}%
\bibitem [{\citenamefont {Komnik}\ \emph {et~al.}(2004)\citenamefont {Komnik},
  \citenamefont {Harting},\ and\ \citenamefont {Herrmann}}]{KHH04}%
  \BibitemOpen
  \bibfield  {author} {\bibinfo {author} {\bibfnamefont {A.}~\bibnamefont
  {Komnik}}, \bibinfo {author} {\bibfnamefont {J.}~\bibnamefont {Harting}}, \
  and\ \bibinfo {author} {\bibfnamefont {H.~J.}\ \bibnamefont {Herrmann}},\
  }\href@noop {} {\bibfield  {journal} {\bibinfo  {journal} {Journal of
  Statistical Mechanics: theory and experiment}\ }\textbf {\bibinfo {volume}
  {P12003}} (\bibinfo {year} {2004})}\BibitemShut {NoStop}%
\bibitem [{\citenamefont {Nguyen}\ and\ \citenamefont
  {Ladd}(2002)}]{nguyen2002lubrication}%
  \BibitemOpen
  \bibfield  {author} {\bibinfo {author} {\bibfnamefont {N.-Q.}\ \bibnamefont
  {Nguyen}}\ and\ \bibinfo {author} {\bibfnamefont {A.}~\bibnamefont {Ladd}},\
  }\href@noop {} {\bibfield  {journal} {\bibinfo  {journal} {Physical Review
  E}\ }\textbf {\bibinfo {volume} {66}},\ \bibinfo {pages} {046708} (\bibinfo
  {year} {2002})}\BibitemShut {NoStop}%
\end{thebibliography}
\end{document}